\documentclass[preprint2]{aastex}
\usepackage{amssymb}
\usepackage{natbib}

\shorttitle{Stochastic acceleration}
\shortauthors{Virtanen & Vainio}

\begin{document}

\title{Stochastic Acceleration in Relativistic Parallel Shocks}

\author{Joni J.P. Virtanen}
\affil{Tuorla Observatory\altaffilmark{1}\altaffiltext{1}%
  {Tuorla Observatory is part of the V\"ais\"al\"a Institute 
    for Space Physics and Astronomy}}
\affil{V\"ais\"al\"antie 20, FI-21500 Piikki\"o, Finland}
\email{joni.virtanen@utu.fi}
\and
\author{Rami Vainio}
\affil{Department of Physical Sciences}
\affil{P.O. Box 64, FI-00014 University of Helsinki, Finland}
\email{rami.vainio@helsinki.fi}

\begin{abstract}
We present results of test-particle simulations on both the first and
the second order Fermi acceleration (i.e., stochastic acceleration) at
relativistic parallel shock waves. We consider two scenarios for
particle injection: (i)~particles injected at the shock front, then
accelerated at the shock by the first order mechanism and subsequently
by the stochastic process in the downstream region; and (ii)~particles
injected uniformly throughout the downstream region to the stochastic
process mimicing injection from the thermal pool by cascading
turbulence.  We show that regardless of the injection scenario,
depending on the magnetic field strength, plasma composition, and the
employed turbulence model, the stochastic mechanism can have
considerable effects on the particle spectrum on temporal and spatial
scales too short to be resolved in extragalactic jets. Stochastic
acceleration is shown to be able to produce spectra that are
significantly flatter than the limiting case of $N(E)\propto E^{-1}$
of the first order mechanism. Our study also reveals a possibility of
re-acceleration of the stochastically accelerated spectrum at the
shock, as particles at high energies become more and more mobile as
their mean free path increases with energy.
Our findings suggest that the role of the second order mechanism in
the turbulent downstream of a relativistic shock with respect to the
first order mechanism at the shock front has been underestimated in
the past, and that the second order mechanism may have significant
effects on the form of the particle spectra and its evolution.

\end{abstract}

\section{INTRODUCTION} 
\label{sec:introduction}

For particle acceleration in astrophysical sources, such as jets and
shock waves, two kinds of Fermi acceleration mechanisms have typically
been considered: the first order Fermi acceleration at the shock
fronts, and the second order Fermi acceleration (often referred to as
stochastic acceleration) in the turbulent plasma. The first order
mechanism is well known to produce a power-law particle spectrum
$N(E)\propto E^{-s}$ with spectral index being a function $s =
(r+2)/(r-1)$ of the compression ratio $r$ for non-relativistic shocks
\citep[e.g.,][]{Drury1983} and in relativistic shocks approaching the
value $s\approx2.2$ at the ultra-relativistic limit
\citep[e.g.,][]{KirkEtAl2000,AchterbergEtAl2001,LP2003,VV2003a}, and
it has been successfully applied in explaining the synchrotron
properties of, e.g., some active galactic nuclei (AGN)
\citep[e.g.,][]{TurlerEtAl2000}. For flatter spectra ($s\lesssim2$)
the first order mechanism is not, however, as attractive. Although it
can produce hard spectra with spectral index $s\to1$ depending, for
instance, on the injection model or the scattering center compression
ratio \citep[e.g., \citeauthor{EllisonEtAl1990}
\citeyear{EllisonEtAl1990}; Vainio, Virtanen \& Schlickeiser 2003,
hereafter referred to as][]{VVS}, it is not able to produce spectral
indices flatter than $s=1$. Stochastic acceleration, on the other
hand, has been known to exist and be present in the turbulent
downstream of shocks, but because it works on much longer time scales
than the first order mechanism \citep[e.g.,][]{CS1992,VS1998} it
has been frequently neglected in studies of relativistic particle
acceleration \citep[however, note e.g.,][]{Ostrowski1993}.

The argument of longer acceleration time scale renders the
second order mechanism less important than the first order one, when
the two mechanisms operate concurrently and, thus, the particle
spectrum at the shock front is considered. When discussing non-thermal
particle distributions radiating in astrophysical objects, however, it
is important to acknowledge that the bulk of radiation is emitted by
the particles that have already left the shock front towards the
downstream. Thus, the second order mechanism has a much longer time
available to accelerate the particles than the first order
mechanism. Although one could justify the neglect of stochastic
acceleration when calculating the particle spectrum right at the shock
front, it is not possible to neglect its effect on the spectrum of
radiating particles in astrophysical shock waves in general as these
objects, especially those believed to host relativistic shock
waves, are often spatially unresolved.

In this paper we have studied the possible effects of stochastic
electron acceleration in parallel relativistic shock waves. We
approach the subject via numerical test-particle 
simulations and present our model
including both the first and the second order Fermi acceleration. We
employ the model to study the effect of stochastic acceleration on (i)
particles injected at the shock front and subsequently accelerated by
the first order mechanism and (ii) particles drawn from the heated
(but not shock accelerated) particle population of the downstream
region of the shock. We focus on shock waves that, in addition to
being parallel, have small-to-intermediate Alfv\'enic Mach
numbers. Low Mach-number relativistic shocks could prevail in
magnetically dominated jets that are lighter than their surroundings,
e.g., because of having a pair-plasma composition.

The structure of this paper is as follows. In \S\ref{sec:model} we
present our model, state the limiting assumptions we have made, and
briefly discuss the limitations caused by the assumptions and
simplifications; the description of the numerical code, as well as the
implementations and mathematical details of the underlying physics are
described in Appendix A. In \S\ref{sec:Results} we
present the results achieved using the simulation code. We begin by
showing that when compared to some relevant previous studies -- both
analytical and numerical -- our results are in very good accordance
with those achieved previously by many authors. We then continue
presenting the results for stochastic acceleration in various cases,
and in \S\ref{sec:Discussion} we discuss our results and their
relationship to both the previous studies and possible future
applications, and list the conclusions of our study. 
Also the limitations caused by the test-particle approach are shortly 
discussed.

\section{MODEL} 
\label{sec:model}

In this section we describe the properties of our model in general,
including the employed assumptions and physics related to them. The numerical
Monte-Carlo approach is described in detail in Appendix A, 
where also implementations of the model properties are discussed.

\subsection{Coordinate systems}

Before proceeding into the model, we define the coordinate systems
employed in our study: the frame where the shock front is at rest is
called \emph{the shock frame}. We consider parallel shocks, i.e.,
shocks where the mean magnetic field and the plasma flow are directed
along the shock normal.
The frame where the bulk plasma flow is at rest is called \emph{the
local plasma frame}; this frame moves with the local flow speed, $V$,
with respect to the shock frame. Finally, we consider \emph{the wave
frame}, which moves with the phase speed $V_\phi$ of the plasma waves
with respect to the local plasma frame and, thus, at speed
$(V+V_\phi)/(1+VV_\phi/c^2)$ with respect to the shock frame denoting,
as usual, the speed of light by $c$.  If the scattering centers are
taken to be fluctuations frozen-in to the plasma then the speed of the
waves with respect to the underlying plasma flow is $V_\phi=0$ and the
plasma frame is also the rest frame of the scattering centers.

\subsection{Shock Structure}
\label{sub:shock_structure}

\begin{figure}[t]
  \plotone{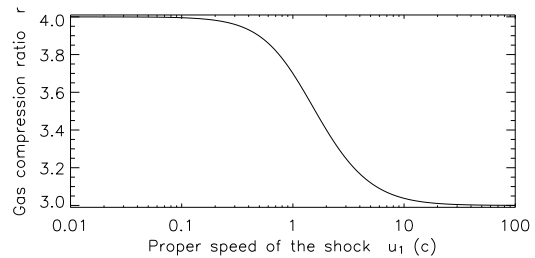}
  \caption{Gas compression ratio $r$ in a parallel shock propagating into
    a cold medium with proper speed $u_1$ \protect\citep{VVS}.}
  \label{fig:compression_ratio}
\end{figure}

We use the hyperbolic tangent profile of Schneider \& Kirk (1989) to
model the flow velocities at different distances from the shock. 
The width of the transition from the far upstream flow speed, $V_{1}$, to
that of the far downstream, $V_{2}$, takes place over a distance of
$0.01\lambda_{\rm e}(\Gamma_1)$ %
(for which the shock can still be considered almost step-like; 
see, e.g., \citealt{VV2003a}), where $\lambda_{\rm e}(\gamma)$
denotes the mean free path of the electrons as a function of Lorentz
factor
\begin{equation}
  \gamma=1/\sqrt{1-(v/c)^{2}},
\end{equation}
$v$ is the electron speed, and $\Gamma_1$ is the Lorentz factor of the
upstream bulk flow. (We use the standard notation of subscript 1[2] 
denoting the upstream [downstream] values.)  The ratio of the flow speeds
on both sides of the shock gives the gas compression ratio
\begin{equation}
  r=\frac{V_{1}}{V_{2}}. \label{eq:compression_ratio}
\end{equation}
We fix the shock speed in the upstream rest frame and, thus, the
(equal) upstream bulk speed $V_{1}$ in the shock frame. Using this we
compute the corresponding gas compression ratio for a given shock
proper speed
\begin{equation}
  u_{1}=c\sqrt{\Gamma_{1}^{2}-1}
\end{equation}
following a scheme described by \citet{VVS} and shown in Figure
\ref{fig:compression_ratio}.  Finally, the flow speed in the far
downstream $V_2$ is given by equation (\ref{eq:compression_ratio}).
%
%

\subsection{Magnetic Field and Scattering} 
\label{sub:magnetic_field}

For modeling the magnetic field structure we adopt the quasilinear
approach, where the field $B$ is considered to consist of two parts:
the static, large scale background field $B_{0}$, and smaller scale
fluctuations with amplitude $\delta B< B_{0}$. We model the turbulence
as being composed of a wide-band spectrum of Alfv\'en waves propagating
both parallel and anti-parallel to the flow. In the local plasma frame
the waves propagate at Alfv\'en speed
\begin{equation}
  v_{\rm A}=\frac{B_{0} c}{\sqrt{4\pi hn + B_{0}^{2}}}  \label{eq:alfven_speed}
\end{equation}
where $h=(\rho+P)/n\equiv w/n$, $n$, $\rho$, and $P$ are the specific
enthalpy, the number density, the total energy density, and the gas
pressure -- all measured in the local plasma frame. The
wave intensity as a function of wavenumber, $I(k)$, is assumed
to have a power-law form for wavenumbers above an inverse correlation length
 $k_{0}.$ We write the power-law as 
\begin{equation}
  I(k)=I_{0}(k_{0}/k)^{q}, \; k>k_0
  \label{eq:fluctuation_spectrum}
\end{equation}
where $q$ is the spectral index of the waves. For wavenumbers smaller
than $k_{0}$ the wave intensity per logarithmic bandwidth is assumed
to be equal to the background field intensity, i.e., $I(k)=B_0^2
k^{-1}$ for $k<k_{0}$. In this work we use two values for $q$: $2$ and
$5/3$. The former produces rigidity independent scattering mean free
paths, while the latter is consistent with the Kolmogorov
phenomenology of turbulence.

Electrons scatter off the magnetic fluctuations resonantly. The
scattering frequency of electrons with Lorentz factor $\gamma$ is
determined by the intensity of waves at the resonant wavenumber (see
Appendix A.)
\begin{equation}
  k_{{\rm res}} = \frac{\Omega_{{\rm e}}}{v}
  = \frac{\Omega_{{\rm e},0}}{c\sqrt{\gamma^2-1}},
  \label{eq:resonant_wavenumber}
\end{equation}
where
\begin{equation}
  \Omega_{{\rm e}}=\frac{\Omega_{{\rm e,0}}}{\gamma}
  \label{eq:gyrofrequency}
\end{equation}
is the relativistic electron gyrofrequency, and $\Omega_{\rm e,0} =
\frac{eB}{m_{{\rm e}}c}$ is its non-relativistic counterpart.

Scatterings are elastic in the wave frame and the existence of waves
propagating in both directions at a given position, thus, leads to
stochastic acceleration \citep{Schlickeiser1989}.  Since the spectrum
of waves is harder below $k=k_0$, scattering at energies
\begin{equation}
  \gamma > \gamma_{0} \equiv \frac{\Omega_{{\rm e,0}}}{k_{0}c}\gg1
  \label{eq:gamma_0}
\end{equation}
becomes less efficient. Thus, we expect the electron acceleration
efficiency to decrease at $\gamma > \gamma_0$. Instead of trying to
fix the value of $k_0$, we use a constant value $\gamma_{0}=10^{6}$,
which corresponds to observations of maximum Lorentz factor of
electrons in some AGN jets \citep{MeisenheimerEtAl1996}.

In addition to scattering, the particles are also assumed to lose
energy via the synchrotron emission. The average rate of energy loss
for an electron with Lorentz factor $\gamma$ in the frame co-moving
with the plasma is given by
\begin{equation}
  \frac{dE}{dt} = -\frac{4}{3}\sigma_{{\rm T}}c\gamma^{2}U_{B},
  \label{eq:s_loss}
\end{equation}
where $\sigma_{{\rm T}}$ is the Thompson cross-section and
\mbox{$U_{B}=B_{0}^{2}/(8\pi)$} is the magnetic field energy
density. We calculate the latter in all simulations by assuming a
hydrogen composition of the plasma.

\subsection{Alfv\'en-Wave Transmission} \label{sub:transmission}

\begin{figure}[t]
  \plotone{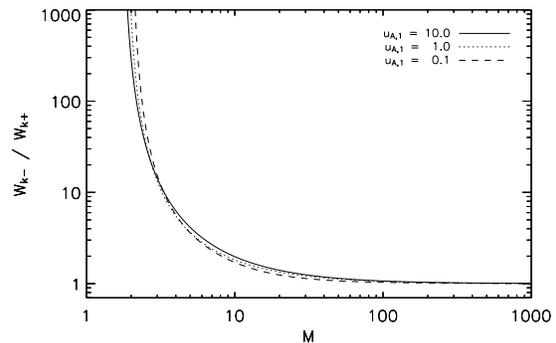}
  \caption{Ratio of the amplified wave intensities as function of
    Alfv\'enic Mach number. Cases corresponding to proper Alfv\'en
    speeds $u_{{\rm A,1}}$ of $10.0\,c$, $1.0\,c$, and $0.1\,c$ are
    plotted with solid, dotted, and dashed lines, respectively, in the
    case of Kolmogorov turbulence, to $q=5/3$.}
  \label{fig:amplification}
\end{figure}

Downstream Alfv\'en-wave intensities can be calculated from know
upstream paramters \citep[e.g.,][]{VS1998,VVS}. Regardless of the
cross helicity of the upstream wave field (only parallel or
anti-parallel waves, or both), there are always both wave modes
present in the downstream region. The transmission coefficients for
the magnetic field intensities of equation
(\ref{eq:fluctuation_spectrum}) of forward $(+)$ and backward $(-)$
waves at constant wavenumber $\tilde{k}$, measured in the wave frame,
can be written \citep[see][equations (22) and (23)]{VVS}, as
\begin{equation}
  \tilde{T}_{\tilde{k}\pm}^{2} \equiv 
  \frac{\tilde{I}_{2}^{\pm}(\tilde{k})}{\tilde{I}_{1}^{\pm}(\tilde{k})}
  \hspace{0.5cm} {\rm and} \hspace{0.5cm}
  \tilde{R}_{\tilde{k}\pm}^{2} \equiv
  \frac{\tilde{I}_{2}^{\mp}(\tilde{k})}{\tilde{I}_{1}^{\pm}(\tilde{k})}.
  \label{eq:T_wk}
\end{equation}
Using these we can solve the amplification factor of the wave intensity
for both wave modes for constant wave frame wave number $\tilde{k}$ as 
\begin{equation}
  \tilde{W}_{\tilde{k}\pm} = 
  \tilde{T}_{\tilde{k}\pm}^{2}+\tilde{R}_{\tilde{k}\mp}^{2}.
  \label{eq:amplification_factor}
\end{equation}
Amplification factor depends on the strength of the magnetic field
as well as on the form of the turbulence spectrum. The intensity 
ratio of the backward waves to that of forward waves as a function 
of the quasi-Newtonian Alfv\'enic Mach number, 
\begin{equation}
  M=u_{1}/u_{{\rm A,1}},
\end{equation}
is plotted in Figure \ref{fig:amplification} for three different Alfv\'en 
proper speed 
\begin{equation}
  u_{{\rm A,1}}=v_{{\rm A,1}}/\sqrt{1-\beta_{{\rm A,1}}^{2}},
\end{equation}
where $\beta_{{\rm A,1}}=v_{{\rm A,1}}/c$. The waves are seen to
propagate predominantly backward for relatively low-Mach-number shocks
as shown by \citet{VVS} in case of relativistic shocks, and by
\citet{VS1998} for non-relativistic shocks. This enables the
scattering center compression ratio $r_{k}$ to grow larger than the
gas compression ratio $r$ and, thus, to cause significantly harder
particle spectra compared to the predictions of theories relying on
fluctuations frozen-in to plasma flow. As the Mach number increases,
the downstream wave intensities approach equipartition at the
ultra-relativistic limit.
%
%

Waves conserve their shock-frame frequencies during the shock crossing
\citep{VS1998}, and for given upstream wave-frame wavenumbers
$\tilde{k}_{1\pm}$ (here equipartition of the upstream waves is
assumed) the downstream wave frame values are obtained from
\citep{VVS}
\begin{equation}
  \tilde{k}_{2+} = \frac{ \Gamma_{1\pm} V_{1\pm} }{ \Gamma_{2+} V_{2+} }
                    \tilde{k}_{1\pm} ; \quad
  \tilde{k}_{2-} = \frac{ \Gamma_{1\pm} V_{1\pm} }{ \Gamma_{2-} V_{2-} }
                    \tilde{k}_{1\pm}
\end{equation}
for both the forward ($\tilde{k}_{2+}$) and the ($\tilde{k}_{2-}$)
waves.  Here $V_{j\pm}$ and $\Gamma_{j\pm}$ refer to the wave speeds
of forward (+) and backward ($-$) waves in the upstream ($j=1$) and
downstream ($j=2$) region and to the respective Lorentz factors.  The
functional form of the spectrum does not change on shock crossing and
the spectral index $q$ in equation \ref{eq:fluctuation_spectrum} is
the same both sides of the shock.

\subsection{Testing the Model}

\begin{figure}[t]
  \plotone{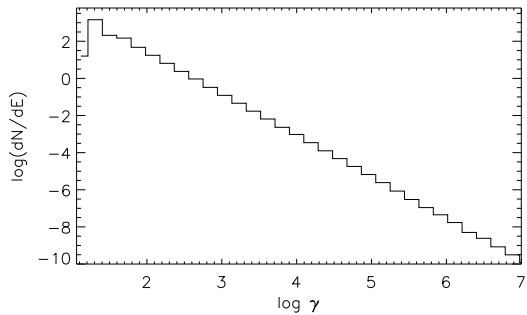}
  \caption{Spectrum of the particles accelerated at the shock front
    in the step-shock approximation. Resulting spectral index 
    for power-law -- extending over five orders of magnitude in energy -- 
    is $s\simeq2.2$. }
  \label{fig:FI-spectrum}
\end{figure}

We have tested the ability of our model to produce results expected
from previous numerical studies and theory. To test the model in the
case of first order acceleration we ran numerous %
test-particle simulations with
different injection energies and shock widths
\citep{VV2003a,VV2003b}. We found our results to be in very good
agreement with the semi-analytical results for modified shocks of
\citet{SK1989}, as well as to the numerical results of, e.g.,
\citet{EllisonEtAl1990} for the corresponding parts of the studies.
For the step-shock approximation, spectra with indices close to the
predicted value of $\sim2.2$ were obtained. An example of the
test-runs is shown in Figure \ref{fig:FI-spectrum} with the shock proper
speed $u_{1}=10\, c$ and compression ratio $r\approx3$, assuming
scattering centers frozen-in into the plasma and turbulence having
a spectral index of $q=2$.

\begin{figure}[t]
  \plotone{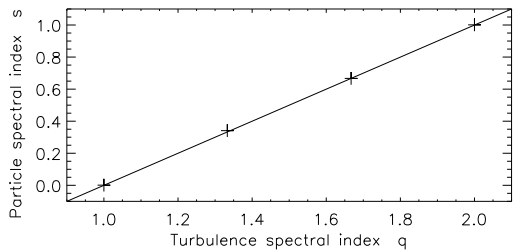}
  \caption{Comparison of the simulation model and the theory. 
    Spectral index of the energy spectrum of particles accelerated 
    stochastically for different spectral indices of the magnetic 
    field fluctuations. Data from the simulations are plotted with crosses 
    (statistical error of each plot is $0.02$ or less), and the
    theoretical prediction of $s=q-1$ is plotted as a straight line.}
  \label{fig:stochastic_test}
\end{figure}

For testing the second order acceleration we chose an analytically
well known case, namely that of uniform flow with waves streaming in
parallel and anti-parallel directions with equal intensities. In this
case the momentum diffusion coefficient of charged particles can be
given as \citep{Schlickeiser1989}
\begin{equation}
  A_{2}(x,p) \simeq \frac{2\pi\Omega_{{\rm e}}^{2-q}k_0^qI_{0}}{q(q+2)B_{0}^{2}}
  \frac{p^{2}v_{{\rm A}}^{2}}{v^{3-q}},
  \label{eq:mom_diff_coefficient}
\end{equation}
which in the relativistic case ($v\approx c$ and $p\gg m_{{\rm e}}c$)
is found to depend on the momentum $p$ as 
\begin{equation}
  A_{2}\propto p^{q}.\label{eq:mom_diff_coeff_relation}
\end{equation}
For a constant particle injection at low momenta, this leads to a
simple relation of the spectral index of the volume-integrated
particle energy spectrum, $s$, and the magnetic field fluctuations,
$q$:
\begin{equation}
  s = q - 1.
  \label{eq:index_relation}
\end{equation}
As a test case, we ran simulations involving only stochastic
acceleration, and found that for values \mbox{$q\in[1,2]$} the model
produces exactly those indices expected from the analytical
calculation. The spectral indices obtained from the simulation are
plotted together with the theoretical prediction in
Figure~\ref{fig:stochastic_test}.
%
%

\section{RESULTS} \label{sec:Results}

In this section we apply our model to stochastic particle acceleration
in the downstream region of a relativistic parallel shock, using a 
test-particle approach.

First we
show how the stochastic process affects a non-thermal particle
population, already accelerated at the shock via the first order
mechanism. Then we consider particles injected throughout the
downstream region. Finally we present an example of the combination of
both injection schemes.  Simulations were run separately for low,
intermediate and high Alfv\'enic-Mach-number shocks ($M=3$, $M=10,$
and $M=1000$, respectively -- see Fig.  \ref{fig:amplification} for
the corresponding wave intensity ratios), and for four cases of
downstream turbulence; for turbulence spectral index $q=2$ and $q=5/3$
with downstream wave field calculated using wave transmission analysis
described in \S \ref{sub:transmission}, and with the downstream forward
and backward waves being in equipartition.%
\footnote{In the case of $M=1000$ the effects were -- at best --
  barely visible, as expected.  For this reason these results are not
  included in this paper, but are available in an electronic form at
  \url{http://www.astro.utu.fi/red/qshock.html}.} The proper speed of
the shock is set to $u_1 = 10\, c$ in all simulations.

\subsection{Electrons Injected and Accelerated at the Shock}

We have studied the effect of stochastic acceleration on particles
that have been already accelerated at the shock. This was done by
injecting particles into the shock and the first order mechanism, and
allowing them to continue accelerating via the stochastic process in
the downstream region. Injection of the particles took place in the
downstream immediately after the shock, and particles were given an
initial energy of a few times the energy of the thermal upstream
particles as seen from the downstream. This kind of injection
simulates some already-energized downstream particles returning into
the shock, but without the need of processing the time consuming bulk
of non-accelerating thermal particles. The high-energy part of the
particle energy distribution -- which we are interested in in this
study -- is similar, regardless of the injection energy.
%
\begin{figure*}
  \begin{center} \begin{minipage}{13.0cm}
  \plotone{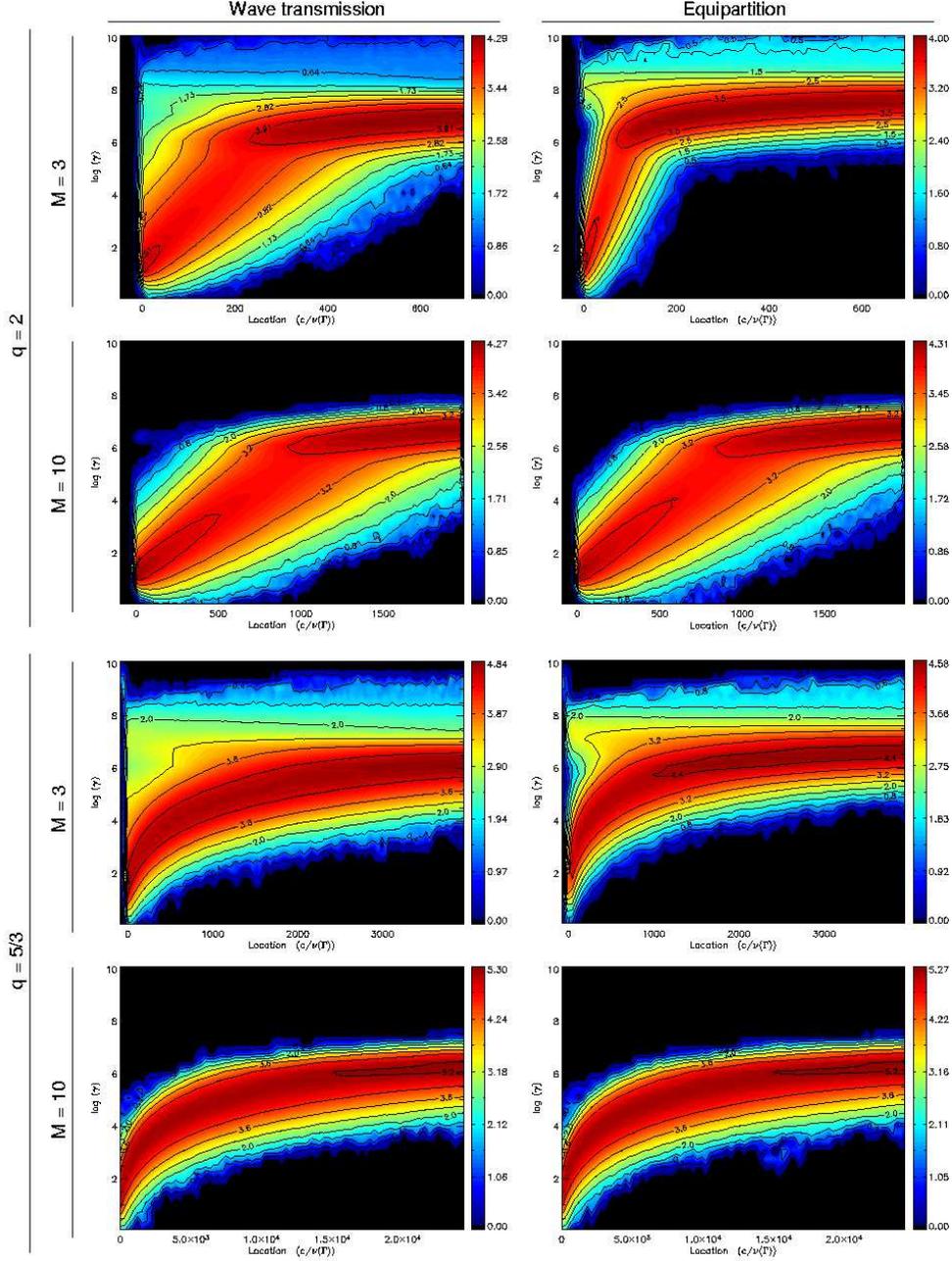}
  \end{minipage}\end{center}
  \caption{Evolution of the particle spectra in the turbulent
    downstream region due to stochastic acceleration. Particles are
    initially injected and accelerated at the shock front, located at $x=0$
    near the left edge of the plots.  Contours of $\log(E\frac{dN}{dE})$
    show the steady-state particle energy distribution.  On the
    left-hand panels the wave transmission calculation of
    \protect\citet{VVS} is applied, whereas on the right-hand panels
    the downstream waves modes are assumed to be in equipartition. In
    the two uppermost rows $q=2$, and in the two lowermost panels
    $q=5/3$. Results for Alfv\'enic Mach numbers $M=3$ (first and
    third row) and $10$ (second and fourth row) are shown for each
    case.  The physical sizes of the downstream region are
    approximately $10^{12}$ cm (first row), and $10^{13}$ cm (second
    row) for the $q=2$ cases, and $10^{11}$ cm (third row), and
    $10^{12}$ cm (fourth row) for the $q=5/3$ cases. }
    \label{fig:distributions_shock}
\end{figure*}
%

We found that in the case of high Alfv\'enic Mach number ($M=1000$,
corresponding to magnetic field $B_0 \simeq 1.4$ mG in a hydrogen
plasma) the contribution of the stochastic process to the energy
distribution of the particles is, indeed, very insignificant compared
to that of the first order acceleration at the shock; the energy
spectrum maintains its shape and energy range unaltered at least for
tens of thousands thermal electron mean free paths.  This is the case
regardless of the applied turbulence spectral index, and because the
analytically calculated wave intensities are very close to
equipartition for high-Mach-number shock (see
Fig. \ref{fig:amplification}), the difference between the analytically
calculated downstream wave field and the explicitely assumed
equipartition case was, as expected, minimal.

For stronger magnetic fields ($M=10$ and $M=3$, corresponding to
\mbox{$0.14$~G} and \mbox{$0.46$ G}, respectively, in a hydrogen
plasma, and to $4.6$~mG and $15$~mG in a pair plasma) the effect of
stochastic acceleration is, on the contrary to the high-$M$ case, very
pronounced. The stochastic process begins to re-accelerate particles
immediately after the shock front, and the whole spectrum slowly
shifts to higher energies as shown in Figure
\ref{fig:distributions_shock}, where the results obtained using wave
transmission analysis are presented in the left-hand column, while the
right-hand panels show the results of the equipartition assumption.
The acceleration rate depends on the wave spectrum in a way that for
$q=2$, for which particles of all energies have the same mean free
path, the stochastic process accelerates particles to higher energies
at constant rate, whereas for the Kolmogorov turbulence the
acceleration rate (like the mean free path of the particles) decreases
as the energy increases, and the energization gradually slows down as
the bulk of the particle energy distribution rises to higher energies.
Also different composition of the downstream wave field leads to
slightly different results: in the ``transmitted'' case (left-hand
panels) the particle population immediately behind the shock front
extends to slightly higher energies than in the equipartition case
(right-hand panels), but for the latter the stochastic acceleration is
clearly quicker. This can be seen best in the uppermost panels of
Figure \ref{fig:distributions_shock}, where the calculated transmitted
wave field differed from the equipartition assumption the most.
%
\begin{figure}[t]
  \plotone{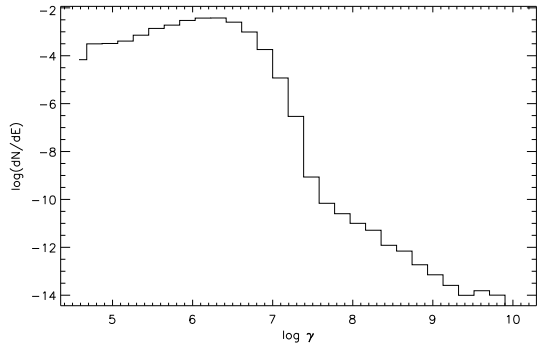}
  \caption{Energy distribution of the particles escaping the shock region 
    and getting absorbed in the downstream.  The Alfv\'enic Mach number of 
    the shock is $M=10$, the proper speed is $u_1 = 10 \, c$, and the 
    turbulence corresponds to $q=5/3$. Note that the distribution is 
    plotted as $\log (dN / dE)$. }
  \label{fig:spectrum_qkM10}
\end{figure}
%

The ``turnover'' of the scattering rate at energy $\gamma_0 = 10^6$
causes the rate of energization to go down for energies greater than
$\gamma_0$. This is because the energy dependence of the mean free
path of the particle changes when particles resonant wavenumber
(eqn. (\ref{eq:resonant_wavenumber})) decreases below $k_0$, which was
use to set the lower limit for the $I(k) \propto k^{-q}$ power-law.
For particles with $\gamma > \gamma_0$ the mean free path starts to
increase much faster, leading not only to the decrease of the
stochastic acceleration efficiency, but also to another notable
effect: as its scattering mean free path increases at
$\gamma>\gamma_0$, the particle will suddenly be able to move much
more easily in the downstream region and even be able to return back
to the shock. Particles already energized first in the shock by the
first order mechanism and then in the downstream by the second order
acceleration, and returning back into the shock again, get
``re-injected'' into the first order acceleration process.  This
effect, in general, is seen in all simulations with $M=3$ as a bending
of the countours to the left at $\gamma>\gamma_0$ close to the shock
(at $x=0$), but it is visible also in $M=10$ shocks in spectra of
particles collected at the downstream free-escape boundary.  An
example of the latter case is shown in Figure
\ref{fig:spectrum_qkM10}.

\subsection{Electrons Injected from the Downstream Bulk Plasma}%
\label{sub:downstream_inj}

\begin{figure*}
  \begin{center} \begin{minipage}{13.0cm}
      \plotone{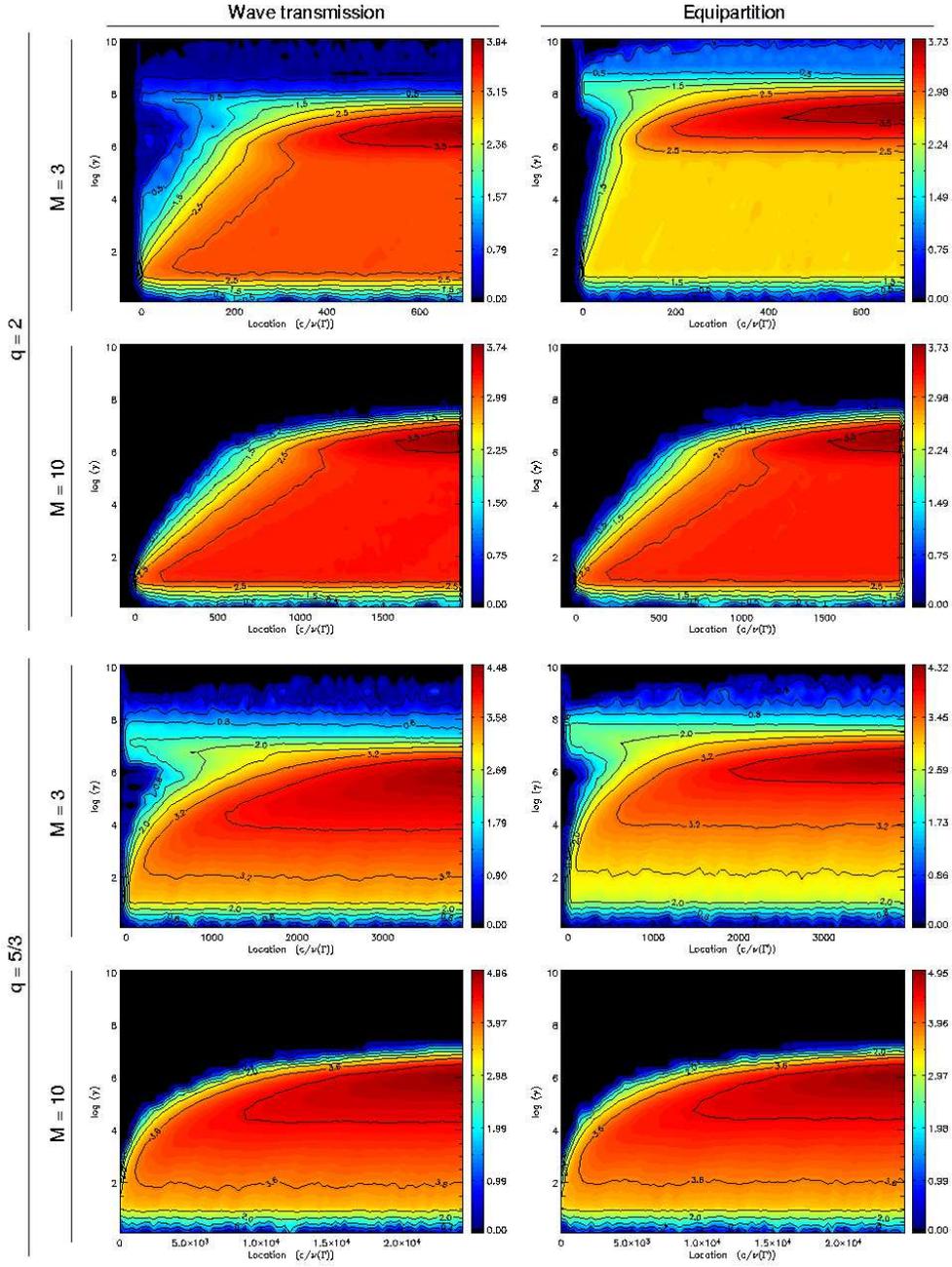}
  \end{minipage} \end{center}
  \caption{Same as Fig. \ref{fig:distributions_shock} but for
    particles injected uniformly and isotropically across the
    downstream region. }
  \label{fig:distributions_downstream}
\end{figure*}

Our second approach was to assume that a constant injection mechanism
exists throughout the downstream region and investigate the stochastic
acceleration process. (Physically, this mimics a case, where turbulent
fluctuations cascade to higher wavenumbers and inject a fraction of
the thermal electrons to the stochastic acceleration process.) We
injected particles at constant energy -- corresponding to the energy
equal to the energy of upstream electrons as seen from the downstream
region -- uniformly and isotropically within the whole downstream
region. The results for different cases are shown in Figure
\ref{fig:distributions_downstream}.
%
%

\begin{figure*}
  \begin{center} \begin{minipage}{0.9\linewidth}
      \plotone{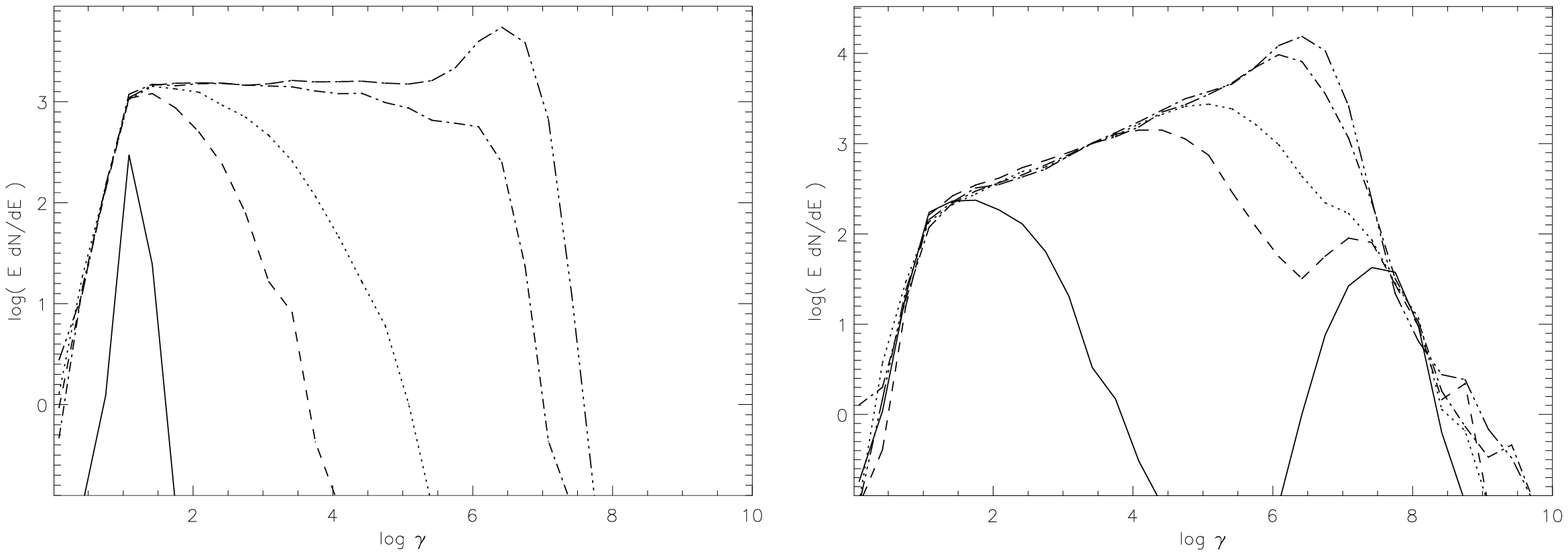}
      \end{minipage} \end{center}
  \caption{Energy spectrum of the particles at different distances
    from the shock.  The particles are injected in the downstream
    region and the distributions are from the simulation with $q=2$
    and $M=10$ (wave field transmitted) for the left-hand panel and
    from $q=5/3$ and $M=3$ (wave field in equipartition) for the
    right-hand panel (corresponding to the left-hand panel of the
    second row and the top right panel of Fig.
    \ref{fig:distributions_downstream}). Slices are from locations
    $x=0$ (solid line), $x=100$ (dashed), $x=400$ (dotted), $x=1000$
    (dash-dotted), and $x=2000$ (dash-dot-dot-dotted) for the
    left-hand panel, and from $x=0$ (solid line), $x=300$ (dashed),
    $x=600$ (dotted), $x=1500$ (dash-dotted), and $x=3000$
    (dash-dot-dot-dotted) for the right-hand panel.}
  \label{fig:slices_downstream}
\end{figure*}

The Figure shows similar behavior as in the case of particles injected
at the shock: significant acceleration was seen only for shocks with
strong magnetic field, while in the $M=1000$ case practically no
acceleration is seen. For the cases with strong magnetic field the
acceleration is seen to work similarly as in the previous
shock-related injection case, energizing particles at constant rate
for $q=2$ turbulence, and with decreasing rate for $q=5/3$. Again the
acceleration rate slows down when energies corresponding to $\gamma_0$
are reached. After this energy particles start to pile up and form a
bump in the distribution immediately behind the $\gamma_0$.  Also here
(at least for $M=3$) some particles with energy $\gamma > \gamma_0$
are able to return to the shock and get re-injected to the first order
process (see, e.g., the right-hand panel of Fig
\ref{fig:slices_downstream}).

In the energy range between the injection energy and $\gamma_0$,
particles begin to form a power-law distribution with spectral index
depending on the magnetic field fluctuations, as expected from
equation (\ref{eq:index_relation}). For $q=2$, the particle spectral
index $s \simeq 1$, and for $q=5/3$, $s \simeq 0.6$. The formation of
the power laws as function of distance from the shock is shown in the
left-hand panel of Figure \ref{fig:slices_downstream} for $q=2$, and
in the right-hand panel for $q=5/3$. In the latter case also the
formation of high-energy bump at the shock by the returning
accelerated particles is seen.
%
%

\subsection{Example of Combination}

We have investigated what kind of particle energy spectra the two
discussed injection schemes -- one operating at the shock, and another
operating uniformly throughout the downstream region -- are able to
create. Next we will present an example of a combination of these. In
the simulation these two cases were kept separate for the sake of
simplicity, but there should be no reason to assume the separation be
present also in nature. Also the relative amounts of shock-injected
and downstream-injected particles is not fixed here, but instead
considered more or less free a parameter.

\begin{figure}[t]
  \plotone{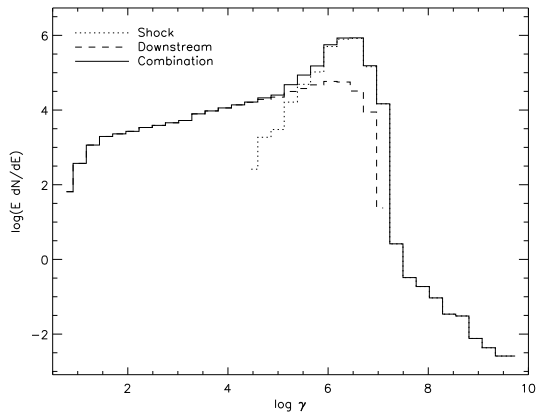}
  \caption{Example of a combination (solid line) energy spectra of
    particles injected to the acceleration process at the shock
    (dashed) and throughout the downstream region (dotted). (Model
    parameters are $q=5/3$ and $M=10$.)  The particles are collected
    at the downstream free escape boundary, $\sim 4\times10^{12}$ cm
    away from the shock, and the number of the particles injected at
    the shock is larger than the number of uniformly injected
    particles by a factor of $100$.}
  \label{fig:qk_M10_combination}
\end{figure}

Assuming different ratios of injected particle populations, the resulting 
spectrum would be very different. An example of combination of the 
two injection schemes in the case of shock with $u_1=10\,c$ and $M=10$, and 
with turbulence corresponding to $q=5/3$, is shown in Figure 
\ref{fig:qk_M10_combination}. 
%
%

\section{DISCUSSION AND CONCLUSIONS}
\label{sec:Discussion}

We have studied stochastic particle acceleration in the downstream
region of a relativistic parallel shock. Applying the wave
transmission calculations of \citet{VVS} and assuming the cross
helicity to vanish in the upstream, we have modeled the turbulence of
the downstream region as a superposition of Alfv\'en waves propagating
parallel and anti-parallel to the plasma flow.  Using a kinetic
Monte-Carlo simulation we have modeled the second order Fermi
acceleration of electrons in the shock environment, and considered
cases of acceleration of downstream-injected particles, as well as
that of particles injected at the shock. We have shown that the
stochastic acceleration can, indeed, have remarkable effects in both
cases. This result is even more pronounced if the two downstream
Alfv\'en wave fields are assumed to be in equipartition.

The behavior of the particle energy distribution in the stochastic
process depends heavily on the strength of the background magnetic
field; in the cases of weak magnetic field and quasi-Newtonian
Alfv\'enic Mach number much larger than the critical Mach number ($M
\gg M_{\rm c} = \sqrt{r}$) the effects of stochastic acceleration are
faded out by the much stronger first order acceleration. Also the
magnetic field turbulence spectrum affects the acceleration
efficiency: for Kolmogorov turbulence with $q=5/3$ the spatial scales
are up to an order of magnitude shorter than in the case of $q=2$
turbulence.  Although the spatial scales in simulations presented here
are enormous compared to those associated with shock acceleration (the
first order process in the immediate vicinity of the shock front), in
case of blazars and other AGNi the scales are still orders of
magnitude too small to be resolved even in the VLBI pictures --
regardless of the turbulence and used magnetic field strength. Also
the acceleration time scales are very short: the time required to
shift the whole spectrum from the initial energy range to $\gamma_{\rm
bulk} \gtrsim 10^6$ ranged from $10$ to $50$ minutes in the $M=10$
case, and for $M=3$ the times were $\lesssim 1$ minute, as measured in
the shock frame.

In addition to the magnetic field strength and turbulence, also the
composition of the downstream wave field was seen to affect the
resulting particle population. When comparing otherwise similar cases
that differ only for the downstream cross helicity (i.e., whether the
wave field is resulting from the wave transmission calculations of
\citet{VVS} or an equipartition of parallel and anti-parallel waves is
assumed), the calculated wave-transmission cases with more
anti-parallel waves \citep[][and Fig.\ref{fig:amplification}]{VVS}
showed stronger first order acceleration, but weaker stochastic
acceleration.  This is because of the larger scattering center
compression ratio in the wave-transmission case leading to more
efficient first order acceleration \citep{VVS} and, on the other
hand, faster momentum diffusion rate in the equipartition case leading
to more efficient stochastic acceleration.

In the cases where the stochastic acceleration was quick enough for
particles to reach the $\gamma_0$ energy while being still
sufficiently close to the shock in order to be able to make their way
back to the upstream region due to their prolonged mean free path, the
first order mechanism was able to re-accelerate the returning
high-energy particles to even higher energies. This led to forming of
a new (quasi-)power-law at energies $\gamma \gtrsim 10^7$ in some
cases.

One notable feature of the present model is that in the case of a
uniform injection process in the downstream region, power-law spectra
with high and low energy cut-offs are formed. Depending on the
turbulence, particle energy spectra have power-law spectral indices of
$0.5$--$1$ with lower and higher energy cut-offs at $\gamma_1 \approx
10^1 \simeq \gamma_{\rm injection}$ and $\gamma_2 \sim 10^6 =
\gamma_0$, respectively. These particles would produce synchrotron
spectra with photon spectral indices $-0.5 < \alpha < 0$ in the
GHz--THz regime for various initial parameters. These properties are
quite similar to those of flat-spectrum sources, for which typical
spectra with $\alpha \gtrsim -0.5$ in the GHz region and flare spectra
with $\alpha \approx -0.2$ in the optically thin region of the
spectrum are seen \citep[e.g.,][and references
therein]{ValtaojaEtAl1988}.  

Combining the resulting energy distributions of both the particles
injected at the shock and the particles injected uniformly in the
downstream region can lead to very different results depending on the
relative amounts of particles injected in both cases. Because
different distributions produce various observable spectra, one might,
in principle, be able to set constraints on different injection
mechanisms, as well as the physical size of the turbulent downstream
region of the shock, comparing the evolution of observed radiated
spectra and the predictions basing on several composite
distributions. Especially the maximum Lorentz factor of electrons
could be used to set limits for $\gamma_0$, as well as the
lowest-energy parts of the non-thermal power-law spectra could give
hints of the injection process.

Our simulations are based on test-particle approximation, 
i.e., the effects of the particles on the turbulent wave spectrum 
and on the shock structure are neglected. 
Including wave-particle interactions in a self-consistent
manner may modify the cross-helicities and wave intensities in the
downstream region and lead to notable effects on the accelerated
particle spectrum \citep[e.g.,][]{Vainio2001}. One should note,
however, that the wave particle interactions are competing with
wave-wave interactions in the turbulent downstream region, which may
modify the the turbulence parameters in a different manner. Including
these effects to our model are, however, beyond of the scope of
the present simulations.

To conclude, the main results of this paper are: 
(i) Stochastic acceleration can be a very efficient mechanism in the
downstream region of parallel relativistic shocks, provided that the
magnetic field strength is large enough in order to make the
Alfv\'enic Mach number approach the critical Mach number ($M_{\rm c} =
\sqrt{r}$) of the shock, i.e., to increase the downstream Alfv\'en
speed enough to allow for sufficient difference in speeds of parallel
and anti-parallel Alfv\'en waves required for rapid stochastic
acceleration.
(ii) The forming of a power-law with very hard particle energy spectra
between the injection energy and $\gamma_0$ in the case of a
continuous injection mechanism in the downstream region. The produced
particle populations could produce synchrotron spectra very similar to
those of flat spectrum sources.
(iii) The interplay between the first and second order Fermi
acceleration at relativistic shocks can produce a variety of spectral
forms not limited to single power laws.

\acknowledgements
J.J.P.V. thanks the Finnish Cultural Foun\-dation for financial support.


\section*{Appendix A.  The Monte Carlo Code}

In this appendix we review the structure and implementation of our
simulation code. The code employs a kinetic test-particle approach; it
follows individual particles in a pre-defined and simplified shock
environment, based on the assumptions and simplifications presented in
\S~\ref{sec:model}. In short, the simulation works as follows: we
trace test particles under the guiding center approximation in a
homogeneous background magnetic field with superposed magnetic
fluctuations. The fluctuations are assumed to be either static
disturbances frozen-in into the plasma flow, or Alfv\'en waves
propagating along the large-scale magnetic field parallel and
antiparallel to the flow (in this paper we apply only the Alfv\'en
wave case). In each time-step the particle is moved and scattered;
scatterings are modeled as pitch-angle diffusion with an isotropic
diffusion coefficient $D_{\mu\mu}\propto1-\mu^{2}$, where
$\mu=\cos\theta$ is the pitch-angle cosine of the particle.
Also for each simulation time-step the energy losses due to the 
synchrotron emission (see eq. (\ref{eq:s_loss})) are calculated.

When the particle passes a free-escape boundary in the far downstream
region it is removed from the simulation. The injection of the
particles into the simulation is modeled using several methods. The
first method involves a uniform and isotropic injection of particles
within one mean free path of a thermal electron downstream of the
shock front, simulating the already-energized supra-thermal particles
crossing and re-crossing the shock.  This injection method allows us
to concentrate on the non-thermal particles without having to spend
most of the computing time simulating the thermal body of the total
particle distribution. Other injection employed methods include an
injection of a cold distribution upstream, or an injection of
particles at thermal energies uniformly and isotropically in the
downstream region (the latter case is applied in \S
\ref{sub:downstream_inj}).

The time unit of the simulation is chosen to be the inverse of the
scattering frequency of an electron having energy $E_{\Gamma_1} \equiv
\Gamma_1 m_{\rm e}c^2$, where $\Gamma_{1}$ is the Lorentz factor of
the shock. The unit of velocity is chosen as $c$. With these choices
the unit of length equals the mean free path of electron with Lorentz
factor that of the shock: $\lambda_{\rm e}(\Gamma_1) = 1 / \nu_{\rm
e}(\Gamma_1)$. 
The time-step is chosen so that it is a small fraction of the inverse 
scattering time, $\Delta t = a \nu_{\rm e}(\gamma)$, where 
$a \lesssim 0.01$.

For a typical simulation $\sim 10^5$ particles are injected. The
number of high-energy particles is further increased by applying a
splitting technique; if the energy of a particle exceeds some
pre-defined value, the particle is replaced by two ''daughter
particles'' which are otherwise identical to their ''mother'', but
have their statistical weight halved. The number of these splitting
boundaries is chosen so that for each simulation the balance between
the statistics and the simulation time is optimal.

\subsection*{A.1. The Shock and the Flow Profile}

We consider a shock wave propagating at speed $V_{1}$ into a cold
ambient medium, which is taken to be initially at rest in the
observer's frame.  In the shock frame the shock is, by definition, at
rest and the upstream plasma flows in with speed $V_{1}$. At the shock
front the plasma slows down, undergoes compression, and flows out at
downstream flow speed $V_2 < V_1$. The shocked downstream values of
the plasma parameters are determined by the compression ratio of the
shock defined in equation (\ref{eq:compression_ratio}).

For the flow speed depending on the location $x$, we use the hyperbolic
tangent form of \cite*{SK1989}, 
\begin{equation}
  V(x) = V_{1} - \frac{V_{1}-V_{2}}{2} \left[ 1 + \tanh \left(
    \frac{x}{W\lambda_{\rm e}(\Gamma_1)} \right) \right],
  \label{eq:FlowProfile_original}
\end{equation}
and fix the shock thickness to $W=0.01$, corresponding to a nearly 
step-like shock. For thicker shocks the first order acceleration 
efficiency drops 
fast \citep[e.g.][]{SK1989,VV2003b}, and also the simple wave-transmission 
calculations of \citet{VVS} would not be valid.

\subsection*{A.2. Magnetic Field}

The strength of the parallel magnetic field, $B_{0}$, in the
simulation is defined with respect to the ''critical strength'' for
which the Alfv\'en speed in the downstream region becomes equal to the
local flow speed; for fields stronger than this the parallel shock
becomes non-evolutionary.  The critical field, $B_{\rm c}$, is
calculated from equation
\begin{equation}
  B_{\rm c}c / \sqrt{4\pi h_{2}n_{2}+B_{c}^{2}} = V_{2},
\end{equation}
for which we get the downstream particle density $n_{2}$ and the
specific enthalpy $h_{2}$ using the magnetohydrodynamical jump
conditions \cite[e.g.,][]{KD1999} and the upstream values $n_1 = 1\
{\rm cm^{-3}}$ and $h_{1}=(\rho_{1}+P_{1})/n_{1}=mc^2$ with $m=m_{\rm
e}+m_{\rm p}$ in a hydrogen plasma and $m=2m_{\rm e}$ in a pair plasma
(throughout this work the upstream plasma is assumed to be cold so
that its pressure may be neglected compared to its rest
energy). Applying the equations for conservation of both energy
\begin{equation}
  h_{1}\Gamma_{1}V_{1}=h_{2}\Gamma_{2}V_{2}
\end{equation}
and mass
\begin{equation}
  \Gamma_{1}V_{1}n_{1}=\Gamma_{2}V_{2}n_{2},
\end{equation}
and a bit of straightforward algebra, the critical field can be written as
\begin{equation}
  B_{\rm c}=\sqrt{4\pi V_{1}V_{2}mn_{1}}\,\Gamma_{1}.
\end{equation}
For example, for a shock with compression ratio $r=3$ propagating into
a cold ambient medium of $n_{1}=1\,{\rm cm^{-3}}$ at proper speed
$u_{1}=10\,c$ the critical field is $B_{\rm c}\approx0.8\,{\rm G}$ for
a hydrogen plasma and $0.03$~G for a pair plasma.

\subsection*{A.3.  Particle Transport and Scattering}

\begin{figure}[t]
  \plotone{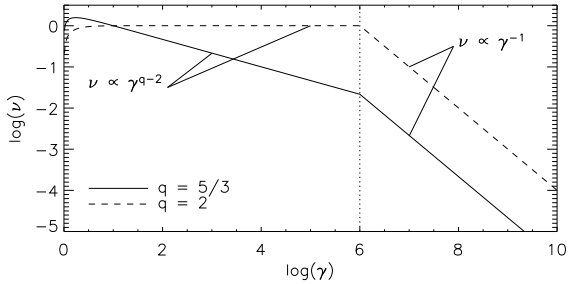}
  \caption{Scattering frequency as a function of Lorentz factor. The
    turn-over energy (eq. (\ref{eq:gamma_0}), see text for details)
    $\gamma_{0}=10^{6}$ is marked with a vertical dotted line. The
    values of $\nu(\mu)$ corresponding to $q=5/3$ and $q=2$ are
    plotted with solid and dashed curves, respectively.}
  \label{fig:scattering-frequency}
\end{figure}

During each Monte Carlo time-step scatterings off the magnetic
fluctuations are simulated by making small random displacements of the
tip of the particle's momentum vector, and the particle is transported
according to its parallel (to the flow) speed in the fixed shock
frame.  In the case of stochastic acceleration, the particle is
scattered twice for each Monte Carlo time-step. Instead of only one
scattering off scattering centers frozen-in into the plasma, the
scattering process is now divided into two parts: the particle is
first scattered in the forward-wave frame, and then immediately again
in the backward-wave frame. The scattering frequencies of both
scatterings are adjusted so that the total statistical effect of the
duplex scattering is consistent with the value of the momentary
scattering mean free path. This is done by balancing the both
scattering frequencies by weight values $\omega_{\pm}\in[0,1]$, which
are calculated from the ratio of the amplified waves (eq.
\ref{eq:amplification_factor} and Fig. \ref{fig:amplification}), so
that they satisfy the relation $\omega_{+}+\omega_{-}=1$. %
Neglecting, for simplicity of the simulations, the dependence of the
scattering frequency on the particle's propagation direction, and
assuming that the scattering is elastic in the rest frame of the
scattering centers, we can use the quasilinear theory: the scattering
frequency of an electron as a function of its Lorentz factor (in the
rest frame of the scattering centers, denoted here by prime) $\gamma'$
is
\begin{equation}
  \nu(\gamma') \approx \frac{\pi}{2}\frac{\Omega_{0}}{\gamma'}
  \frac{k'I'(k')}{B^{2}}  \propto 
  \frac{\left(\gamma'^{2}-1\right)^{(q-1)/2}}{\gamma'}.
  \label{eq:ScatteringFrequency}
\end{equation}
This equation is applicable to particles scattering off waves with
wavenumber larger than $k_{0}$; for particles with Lorentz factor
$\gamma' > \gamma'_{0}$ the scattering frequency decreases as
$\nu\propto\gamma^{-1}$, as at these wavenumbers $k'I(k')=B_0^2$ is
assumed. Scattering frequency as a function of particle's Lorentz
factor is plotted in Figure \ref{fig:scattering-frequency} for
magnetic field fluctuations with $q=2$ (leading to energy-independent
mean free path) and $q=5/3$.
%
%

\begin{figure}[t]
  \plotone{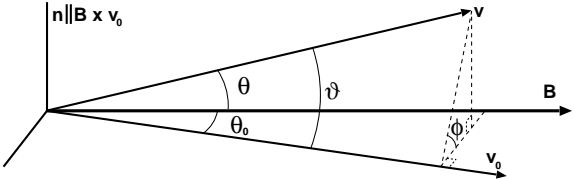}
  \caption{The geometry of the scattering event. Particle is initially
    moving at velocity $\mathbf{v}_{0}$; the angle between the
    velocity and magnetic field $\mathbf{B}$ is $\theta_{0}$. (Thus,
    $\theta_0=\arccos\mu_0'$.) Scattering changes the direction of the
    particle by an angle $\vartheta$ so that the new velocity
    $\mathbf{v}$ has an angle $\theta$ between it and the magnetic
    field.  $\phi$ is the azimuth angle of the new velocity measured
    from the plane defined by $\mathbf{B}$ and $\mathbf{v}_{0}$.}
  \label{fig:geom_of_the_scattering}
\end{figure}

In each scattering, the velocity vector of the particle is first
Lorentz-transformed into the frame of the scatterers, then the new
pitch-angle cosine (in the scatterer frame) is computed from the
formula \citep[e.g.,][]{EllisonEtAl1990}
\begin{equation}
  \mu' \gets \mu'\cos\vartheta+\sqrt{1-\mu'^{2}}\sin\vartheta\cos\phi,
\end{equation}
where the angle between the velocities before ($\mathbf{v_0}$) and
after ($\mathbf{v}$) the scattering, $\vartheta\in[0,\pi)$, and the
angle measured around the scattering axis, $\phi\in[0,2\pi)$ (angles
and the geometry is sketched in
Fig. \ref{fig:geom_of_the_scattering}), are picked via a random
generator from exponential and uniform distributions, respectively
\citep[see][for details]{VainioEtAl2000}. The new velocity vector is
then Lorentz transformed back to the shock frame, and finally the the
particle is moved according to its new parallel velocity.
%
%


\bibliographystyle{apj}

\end{document}